# Particle cooling in Vaporizing Coolant

Ivan V. Kazachkov[1,2]

Dept of Information Technologies and Data Analysis, Nizhyn Gogol state university, Ukraine
Dept of Energy Technology, Royal Institute of Technology (KTH), Stockholm, Sweden)
Ivan.Kazachkov@energy.kth.se

**Abstract:** The approximate mathematical model for a cooling of the particle in a volatile liquid is developed and analyzed. Despite the precise model is complex and requires the solution of the nonstationary two-phase flow equations with the conjugated heat transfer boundary problem for the particle, vapor, and liquid cooling pool, the considered simple model may be of interest. Vapor is permanently produced and removed from the coolant's pool. Analysis of the model obtained resulted in some correlations for the three main parameters of the cooling process, which may be used for estimation of the particle's cooling.

## 1. Introduction and statement of the problem

In many industrial and technical applications (granulation of metal and allows, development of the amorphous materials, etc.) [1-18] the problem of cooling of the high-temperature drops and particles is of great importance. For example, by the development of the passive protection systems against severe accidents at the nuclear power plant, one of the key problems is cooling the corium melt, which is speaded in a containment [19]. By cooling of the high-temperature particles and drops in a volatile coolant, an intensive vaporization around the particle (crisis of heat transfer) is dramatically decreasing the intensity of cooling. Vapor around cooling particle decrease heat transfer by 2-3 decimal orders, which is a so-called crisis of heat transferThe theory and methods of parametric excitation and suppression of oscillations on the film flow surfaces under an action of the electromagnetic fields and vibrations.

Experimental facilities for rapid cooling of the drops, with up to ten thousand degrees per second [3, 13-15], which allow producing amorphous nanostructured granules for the new materials' production with unique properties are of serious interest for application and as such must be developed further. Therefore, the presented paper is devoted to studying of the particle cooling in a vaporizing coolant. The theoretical base and engineering applications, including the new methods and devices for granulation of the liquid metals, have been developed at the Institute of Electrodynamics of the Ukrainian National Academy of Sciences [1-15]. The principle of operation of the installations for electromagnetic batching of metal alloys with a melting temperature up to $10^3$ Celsius degrees and slightly over, which intended for production of the metallic granules, is based on the jet disintegration on the first mode of Eigen oscillations. Applying the crossed electric and magnetic fields at the nozzle leads to a resonance regime with a frequency of the alternating ponderomotive electromagnetic force equal to a frequency of the jet's Eigen oscillations). The magnetic field is created by a permanent magnet. A current in a melt at the nozzle has the industrial frequency 50 Hz.

The installation includes such elements as an induction furnace, a liquid-metal pump for delivering a melt through a channel to the electromagnetic batcher, and a crystallizer of the drops (for obtaining the granules). Electromagnetic batcher implements the controlling capillary disintegration (fragmentation) of a cylindrical jet of liquid metal in the range of 100...400 Hz (electromagnetic force has 100 Hz frequency, it is created applying the electric current of 50 Hz). The diameter of metal particles (granules) is determined by the resonant frequency of a jet at the first harmonics, which is approximately $\omega=0.23u/d$ for the low viscous melts [2, 10-12]. The velocity of liquid metal flow $u$ and the nozzle diameter $d$ determine the size of particles due to the disintegration of a jet and productivity of the granulation machine. A diameter of the spherical particles, which are produced due to a jet's decay, is roughly estimated as $d_p=1.88d$. Because the capillary forces are growing with a decrease of the nozzle's diameter (inversely proportional), the controlled disintegration of a jet is available up to the nozzle's diameter of about 1 mm. Afterward the Coanda effect makes the jet chaotically directed flow from the nozzle. It is breaking the jet's strictly vertical flow. In addition, even an accidental slight clogging of the melt with some impurities can also destroy the jet's flow regime and cause the granulator to stop. Therefore, the jet's type granulators have a restriction in a size of the resulting particles about 2 mm in diameter.

Because the jet granulators, despite having big advantage in getting nearly ideal spherical granules of the same given size, revealed serious impediment for the size limitation about 2 mm from below, the theory of parametrically controlled disintegration of the liquid metal film flow has been developed [1, 3, 13-15], and the methods and devices for film granulation were invented [5-9]. The last ones produce granules in the range of 0.5-1.5 from the average size, but they practically do not have any limitations on a size of the producing granules up to the micrometer size. Due to this, the high cooling rate was obtained in a liquid nitrogen (up to $10^4$ Celsius degrees per second). For this, except the small size of the drops (granules after their solidification), the avoiding of heat transfer crisis was invented. Drops and particles during their cooling were passing through a series of the liquid metal films to permanently destroy the vapor film on the hot particle in a liquid coolant.

Examples of the processes for obtaining the liquid drops of given size with their further solidification into granules are presented in Fig. 1 and Fig. 2 (the drops and particles were cooled in the film granulation machine with high cooling rate up to $10^4$ ºC):

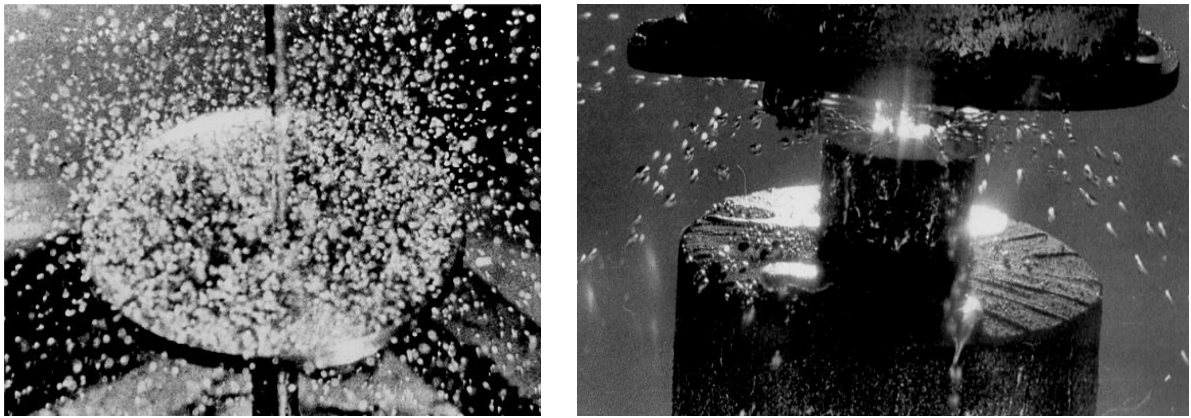

Fig. 1 Vibration controlled soliton-like regime (to the left) and electromagnetically controlled the resonant disintegration of the film flow (to the right)

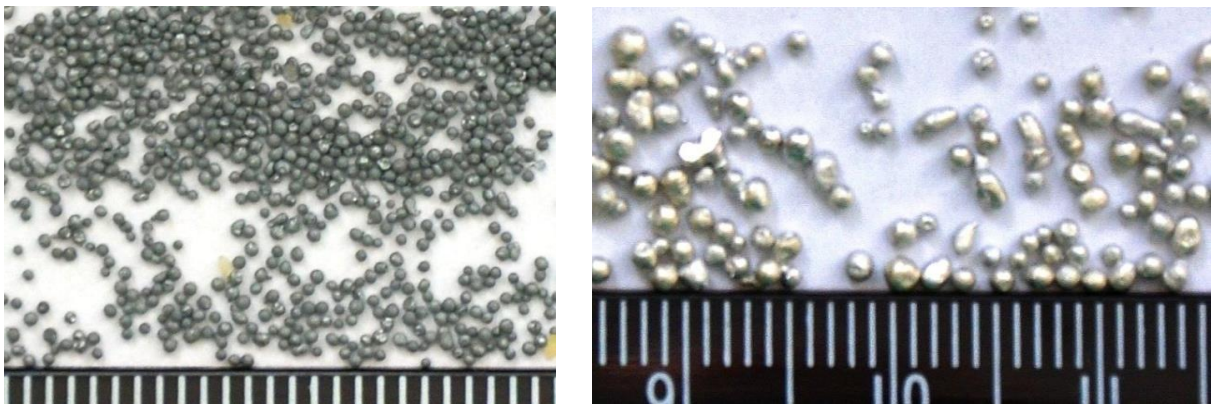

Fig. 2 Granules of the zink (left) and aluminium (right) cooled in a liquid nitrogen

## 2. Physical statement for the model of the particle cooling

Let us consider physical situation according to the pictures presented in Fig. 3. The hot particle is surrounded by a vapor, which is intensively flowing around a particle. Vapor is permanently produced due to intensive heat release from the hot particle. In a granulation machine, it is around thousand Celsius degrees or so, while during severe accidents at the nuclear power plants (NPP) the drops and particles of the corium may be of temperature even higher (2-3 thousand Celsius degrees) [19]. In case of granulation the phenomenon of the crisis of heat transfer substantially worsens the coolability conditions and the cooling rate. In case of cooling of the corium melt during postulated severe accidents at NPP this phenomenon is crucial as far as permanently generating heat by the radioactive nuclear fuel must be removed with the 100% guaranty to avoid the catastrophe.

The described important heat transfer problem is considered in this paper through development and analysis of the mathematical model for particle cooling.

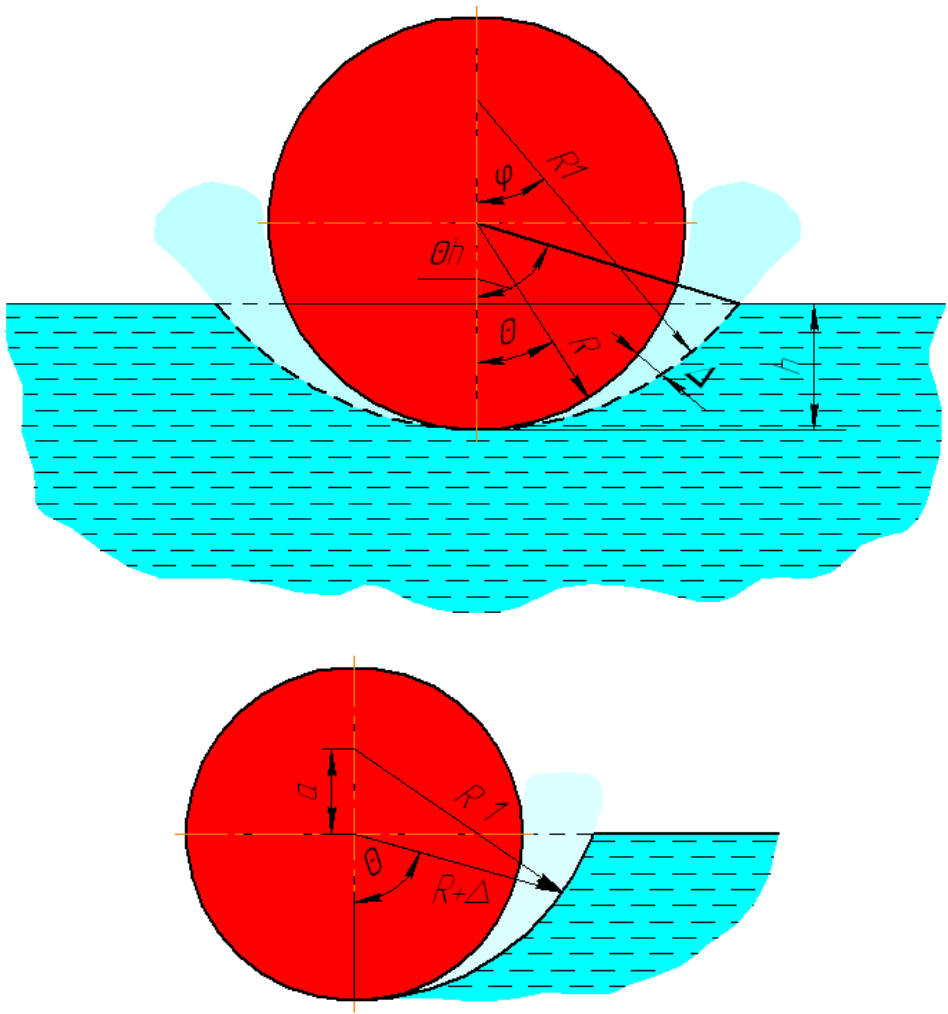

Fig. 3 Schematic of the hot particle in a volatile coolant with a vapor flow around the particle

## 3. A mathematical model for the cooling process of a particle in a volatile liquid

### *Mechanical equilibrium of a particle on a surface of liquid coolant*

Development of the mathematical model for physical processes during the cooling of a drop or particle is done starting from the mechanical equilibrium of particle on a surface of liquid coolant (e.g. in case of the drop falling down into a pool of coolant):

$$\rho_p g \frac{4}{3}\pi R^3 = 2\pi \int_0^{\theta_h} R\sin\theta (p_v \cos\theta + \tau_w \sin\theta) R d\theta, \qquad (1)$$

where

$$p_v = p_0 + \rho_l g\left[R_1 - (R - R\cos\theta)\right] + \frac{2\sigma}{R_1}, \qquad (2)$$

here $R_1$ is the radius of the equivalent circle of the capillary meniscus, $\rho_p, R, g$ are the density and radius of particle and acceleration due to gravity, correspondingly, $\rho_l$ - density of liquid coolant, $\sigma$ - the coefficient of a surface tension; $\theta, \varphi$ are the angles in the considered spherical coordinate system, $p_v, \tau_w$ are the pressure of vapor and the shear stress due to a vapor flow around the particle, respectively, $\theta_h$ is the angle corresponding to a length $h$ of a particle penetration into a coolant. The parameter $\Delta(\theta)$ in Fig. 3 is a width of a layer of vapor flow. The function $\Delta(\theta)$ and a length $h$ of penetration into a coolant depend on particle and coolant density ratio, the temperature of the particle and other physical properties, which are revealed later on. Those determine the intensity of the coolant's vaporization, which, in turn, determines the action on a particle hydrodynamic force.

The equation (1) represents the condition of a particle levitation on the surface of liquid coolant. The weight of the particle is in the equilibrium with two hydrodynamic forces acting on the particle: pressure and shear force. Therefore, the length $h$ of particle's penetration into a coolant may be different depending on the above-mentioned physical conditions.

The radius $R_1$ of the equivalent circle of the capillary meniscus is connected to the width of a vapor layer $\Delta$ as follows:

$$R_1^2 = a^2 + (R+\Delta)^2 - 2a(R+\Delta)\cos(180^0 - \theta) = a^2 + (R+\Delta)^2 + 2a(R+\Delta)\cos\theta,$$

where $R_1 = R + a + \Delta_0$, $a$ is the distance between the center of the particle and the equivalent circle of the capillary meniscus, to be computed further. Here from yields

$$R + \Delta = -a\cos\theta \pm \sqrt{R_1^2 - a^2 \sin^2\theta}.$$

If $\Delta$ is small, then $R_1^2 \gg a^2$, and by small values of $\theta$, which are even smaller ones, is got

$$R + \Delta \approx -a\cos\theta + R_1 - \frac{a^2}{2R_1}\sin^2\theta,$$

and then

$$\Delta = 2a\sin^2\frac{\theta}{2}\left(1 - \frac{a}{R_1}\cos^2\frac{\theta}{2}\right) + \Delta_o. \qquad (3)$$

In the expression (3) thus obtained the second term in the brackets is small comparing to 1, therefore $\Delta \approx 2a\sin^2\frac{\theta}{2} + \Delta_0.$

Now the equation (2) can be represented in a form

$$p_v = p_0 + \rho_l g\left(h - 2R\sin^2\frac{\theta}{2}\right) + \frac{2\sigma}{R_1}. \qquad (2')$$

Then compute the shear stress on a surface of a particle:

$$\tau_w = 3\frac{\mu_v V_v}{\Delta}, \quad \text{or} \quad \tau_w = 3\frac{\mu_v \dot{m}_v}{2\pi\rho_v \Delta^2 R \sin\theta}. \qquad (4)$$

and substituted the ($2'$), (3), (4) into the equation (1), we get the following

$$\frac{2}{3}\rho_p gR = \int_0^{\theta_R}\left\{\left[p_0 + 2\rho_l gR(\sin^2\frac{\theta_R}{2} - \sin^2\frac{\theta}{2}) + \frac{2\sigma}{R_1}\right]\cos\theta + 3\frac{\mu_v \dot{m}_v}{2\pi\rho_v R\Delta^2}\right\}\sin\theta d\theta, \qquad (5)$$

with account $h = 2R\sin^2\frac{\theta_h}{2}$. Here $\mu_v, V_v$ are dynamic viscosity coefficient and velocity of vapor, $\dot{m}_v$ - derivative by time from a mass of vapor (vapor mass rate).

### *Thermodynamic equilibrium of a particle and a coolant*

According to the above-mentioned and the schematic of the model physical situation (Fig. 3), the heat balance for the vapor flow in a layer between the particle and liquid coolant of consideration is as follows (difference of enthalpies between vapor and liquid spent in a vapor flow)

$$(h_v - h_l)d\dot{m}_v = \frac{k_v R^2 \sin\theta}{\Delta}2\pi(T_v - T_l)d\theta, \qquad (6)$$

where from, because of (6) we obtain

$$\Delta = 2\pi R^2 \frac{k_v(T_v - T_l)\sin\theta}{(h_v - h_l)d\dot{m}_v}d\theta. \qquad (7)$$

Here $h_v, h_l$ are enthalpies of the vapor and liquid coolant, respectively, $T_v, T_l$ - temperatures of the vapor and liquid, $k_v$ - the coefficient of the heat conductivity in a vapor flow. The local temperature of a vapor and particle are considered equal due to the small size of the particle and comparably fast local temperature equilibrium process between vapor and particle.

### *Momentum conservation for the elementary volume*

Now the equation of the momentum conservation is considered for the elementary volume of a vapor in a layer between the cooling particle and liquid coolant pool as shown in the Fig. 4:

$$(\tau_w + \tau_l)R^2\sin\theta d\theta + \rho_v gR^2\sin^2\theta\Delta d\theta + dp_v R\Delta\sin\theta = 0, \qquad (8)$$

$$dp_v = -2\rho_l gR \cdot 2\sin\frac{\theta}{2}\cos\frac{\theta}{2}\cdot\frac{1}{2}d\theta = \rho_v gR\sin\theta d\theta, \qquad (9)$$

$$(\tau_w + \tau_l) = \beta\frac{\mu_v V_v}{\Delta} = \frac{\beta\mu_v \dot{m}_v}{2\pi\rho_v 4a^2 \sin^4\frac{\theta}{2}}R\sin\theta. \qquad (10)$$

where $\tau_l$ is a shear stress of a vapor layer with a liquid coolant, $\beta$ is a coefficient (unknown for the moment).

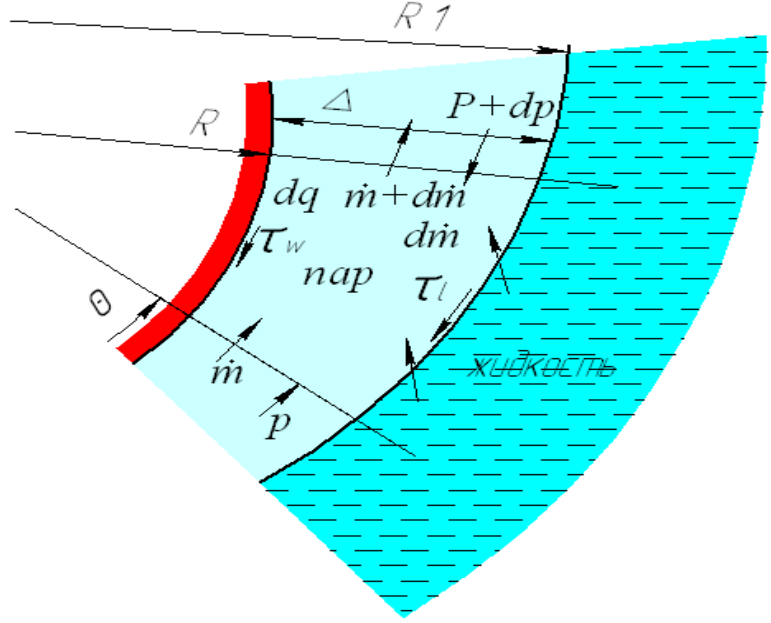

Fig. 4 To the mathematical model of a particle's cooling

With an account of the equations (8) - (10), yields:

$$\frac{\beta\mu_v \dot{m}_v R^2 \sin\theta d\theta}{2\pi\rho_v \Delta^2 R\sin\theta} + \rho_v gR^2 \sin^2\theta\Delta d\theta = \rho_l gR^2 \sin^2\theta\Delta d\theta,$$

where from:

$$\frac{\beta\mu_v \dot{m}_v d\theta}{2\pi\rho_v \Delta^2 R} + \rho_v g\sin^2\theta\cdot\Delta d\theta = \rho_l g\sin^2\theta\cdot\Delta d\theta, \quad \Delta = \frac{\beta^{\frac{1}{3}}\mu_v^{\frac{1}{3}}\dot{m}_v^{\frac{1}{3}}}{[2\pi\rho_v R(\rho_l - \rho_v)g]^{\frac{1}{3}}\sin^{\frac{2}{3}}\theta};$$

which accounting the (6) further results in the following:

$$\beta^{\frac{1}{3}}\mu_v^{\frac{1}{3}}\dot{m}_v^{\frac{1}{3}}d\dot{m}_v = \frac{k_v R^2 \sin\theta\cdot 2\pi(T_v - T_l)}{h_v - h_l}[2\pi\rho_v R(\rho_l - \rho_v)g]^{\frac{1}{3}}\sin^{\frac{2}{3}}\theta d\theta,$$

or

$$\beta^{\frac{1}{3}} \dot{m}_v^{\frac{1}{3}} d\dot{m}_v = 2\pi \left[ \frac{k_v^3 R^7 \rho_v g (\rho_l - \rho_v)(T_v - T_l)^3}{\mu_v (h_v - h_l)^3} \right] \sin^{5/3} \theta d\theta. \quad (11)$$

## 4. Analysis of the mathematical model for a particle in a volatile coolant

### *A solution of the equation*

Now we can perform the integration of the equation (11) thus obtained with the initial condition $\dot{m}_{v,0} = 0$ (zero mass flow rate at the start of the particle's cooling process):

$$\dot{m}_v = \frac{2\pi}{\beta^{1/4}} \left( \frac{3}{4} \int_0^\theta \sin^{2/3} \theta d\theta \right) \left[ \frac{k_v^3 R^3 \rho_v g (\rho_l - \rho_v)(T_v - T_l)^3}{\mu_v (h_v - h_l)^3} \right]^{\frac{1}{4}} R. \quad (12)$$

Substituting the expression (8) into (5) yields the following equation for the function $\theta_h$:

$$\frac{2}{3}\rho_p gR = \left( \rho_0 + \frac{\sigma}{R_1} + 2\rho_l gR \sin^2 \theta_h \right) \frac{1}{2} \sin^2 \theta_h - \rho_l gR \frac{1}{6} \left( \sin^2 \theta_h - 4\sin^2 \frac{\theta_h}{2} \cos^2 \theta_h \right) +$$

$$+ \frac{3\mu_v h_v R(T_v - T_l)}{\rho_v (h_v - h_l) a \Delta_0} \left[ \frac{1 - \ln(1 + \frac{2a}{\Delta_0} \sin^2 \frac{\theta_h}{2})}{1 + \frac{2a}{\Delta_0} \sin^2 \frac{\theta_h}{2}} - 1 \right]. \quad (13)$$

We have two equations for function $\dot{m}_v$ - (10) and (12), and the equation (13) for $\theta_h$. The unknowns here are: the length $h$ of a particle penetration into a coolant (the corresponding angle is $\theta_h$), the initial width of a vapor film on a particle $\Delta_0$, and the radius of an equivalent circle $R_1$ (or $a = R_1 - R - \Delta_0$).

To close the equation array (8), (12), (13), we need to compute

$$\dot{m}_{v,h} = \frac{\alpha_p 2\pi R^2 \sin^2 \frac{\theta_h}{2} (T_w - T_l)}{(h_v - h_l)}, \quad (14)$$

Based on the given values of the parameters $T_w, T_l, h_v, h_l, \theta_h, \alpha_p, R$, where $T_w, \alpha_p$ are the temperature of the particle and heat transfer coefficient from particle to a coolant. Thus, substituting (14) into (10), (12) results in

$$a = \frac{\Delta_0}{2} \left[ \exp\left( \frac{2a}{k_v} \alpha_p \sin^2 \frac{\theta_h}{2} \right) - 1 \right] \sin^{-2} \frac{\theta_h}{2}, \quad (15)$$

$$\frac{\sin^2 \frac{\theta_h}{2}}{\left( \frac{3}{4} \int_0^{\theta_h} \sin^{5/3} \theta d\theta \right)^{3/4}} = \left[ \frac{k_v^3 \rho_v g (\rho_l - \rho_v)(h_v - h_l)^3}{R \beta \mu_v \alpha_p^4 (T_w - T_l)^3} \right]^{\frac{1}{4}}.$$

*Analysis of the solution obtained*

With estimation made for a small length of a particle penetration into the cooling pool, $\sin^2 \frac{\theta_h}{2} \approx \frac{\theta_h^2}{4}$, $\sin^{\frac{2}{3}} \theta \approx \theta^{\frac{2}{3}}$, therefore from (15), (13) follows:

$$\theta_h = 0{,}45 \left(\frac{4}{\alpha_p}\right)^{4/3} k_v \left[\frac{\rho_v g (\rho_l - \rho_v)(h_v - h_l)^3}{R\beta\mu_v(T_w - T_l)^3}\right]^{\frac{1}{3}}, \quad a = \frac{\Delta_0}{2}\left[\exp\left(2\frac{\alpha_p a \theta_h^2}{k_v}\right) - 1\right], \quad (16)$$

$$\frac{2}{3}\rho_p g R = \left(p_0 + \frac{2\sigma}{R + \Delta_0 + a}\right)\frac{\theta_h^2}{2} + \frac{3\mu_v h_v R(T_w - T_l)}{\rho_v (h_v - h_l) a \Delta_0}\left[\frac{1 - \ln(1 + \frac{a\theta_h^2}{2\Delta_0})}{1 + \frac{a\theta_h^2}{2\Delta_0}} - 1\right].$$

Thus, (16) determines three main parameters of the process under investigation: $a, \Delta_0, \theta_h$ for the given physical data. Because of $Nu = \alpha_p a / k_v \sim 10^2 \div 10^3$, and $\theta_h \sim 0{,}2 \div 0{,}5$, we can consider two cases:

$$Nu\theta_h^2 / 2 \ll 1 \quad \text{and} \quad Nu\theta_h^2 / 2 \gg 1. \quad (17)$$

The last case in (17) better corresponds to our conditions, especially for heavy particles, when contact area of the particle with coolant is larger. Therefore:

1) $\theta_h^2 \alpha_p a \ll k_v$, $\quad \theta_h = 0{,}45 \left(\frac{4}{\alpha_p}\right)^{4/3} k_v \left[\frac{\rho_v g (\rho_l - \rho_v)(h_v - h_l)}{R\beta\mu_v(T_w - T_l)}\right]^{\frac{1}{3}}$,

$\Delta_0 = \frac{4k_v}{\alpha_p \theta_h^2}$, $\quad a \ll \Delta_0$, $\quad a = \frac{\Delta_0 \rho_v (h_v - h_l)}{3\mu_v h_v (T_w - T_l)}\left[\left(\rho_0 + \frac{2\sigma}{R + \Delta_0}\right)\frac{\theta_h^2}{2} - \frac{2}{3}\rho_p g R\right];$

2) $\theta_h^2 \alpha_p a \gg k_v$, $\quad \theta_h = 0{,}45 \left(\frac{4}{\alpha_p}\right)^{4/3} k_v \left[\frac{\rho_v g (\rho_l - \rho_v)(h_v - h_l)}{R\beta\mu_v(T_w - T_l)}\right]^{\frac{1}{3}}$,

$\Delta_0 = 2a\exp\left(-\frac{a\alpha_p}{2k_v}\theta_h^2\right)$, $a \gg \Delta_0$: a) $\frac{\theta_h^2}{4}\exp\left(\frac{a\alpha_p}{2k_v}\theta_h^2\right) \ll 1$; б) $\frac{\theta_h^2}{4}\exp\left(\frac{a\alpha_p}{2k_v}\theta_h^2\right) \gg 1$;

$$\left(\frac{2}{3}\rho_p g R - p_0 \frac{\theta_h^2}{2}\right)a^2 + \left[\frac{2}{3}\rho_p g R^2 - p_0 \frac{\theta_h^2}{2}R - \sigma\theta_h^2 + \frac{3R\mu_v h_v (T_w - T_l)}{\Delta_0 \rho_v (h_v - h_l)}\right]a + \frac{3R\mu_v h_v (T_w - T_l)}{\Delta_0 \rho_v (h_v - h_l)}R = 0.$$

Further according to the expressions obtained we can analyze the main parameters of the process for each specific case.

## 5. Conclusion

The developed mathematical model fits for a rough approximation of the particle's cooling in a volatile liquid. The precise model is complex and requires the solution of the nonstationary two-phase flow equations with the conjugate heat transfer problem for the particle, vapor, and liquid cooling pool. Vapor is permanently produced and removed from the coolant's pool. The process is going until the moment when a temperature of the particle during its cooling falls down below the vaporization temperature. Then particle is further cooled by direct contact with the cooling liquid.